\documentclass{article}

\usepackage{PRIMEarxiv}
\usepackage[normalem]{ulem}

\usepackage[T1]{fontenc}    % use 8-bit T1 fonts
\usepackage{hyperref}       % hyperlinks
\usepackage{url}            % simple URL typesetting
\usepackage{booktabs}       % professional-quality tables
\usepackage{amsfonts}       % blackboard math symbols
\usepackage{nicefrac}       % compact symbols for 1/2, etc.
\usepackage{microtype}      % microtypography
\usepackage{lipsum}
\usepackage{fancyhdr}       % header
\usepackage{graphicx}       % graphics
\graphicspath{{media/}}     % organize your images and other figures under media/ folder

\usepackage{amsmath,amssymb,amsfonts}
\usepackage{algorithmic}
\usepackage{algorithm}
\usepackage{graphicx,color}
\usepackage{textcomp}
\usepackage{amsmath}
\usepackage[english]{babel}
\usepackage[utf8x]{inputenc}
\usepackage[T1]{fontenc}
\usepackage{caption}
\usepackage{subcaption}
\usepackage{amssymb}
\usepackage{amsfonts}
\usepackage[font=scriptsize]{caption}
\usepackage{mwe}
\usepackage{mathptmx}
\usepackage[11pt]{moresize}
\usepackage[font=small,skip=0pt]{caption}
\usepackage{float}
\usepackage{caption}
\usepackage{subcaption}
\usepackage{xcolor}
\usepackage{lineno}

\usepackage{tikz}
\usepackage{graphicx}
\usetikzlibrary{positioning}

\usepackage[flushleft]{threeparttable}
\usepackage{calc}
\usepackage{booktabs}
\usepackage{tabularx}% (for comparison)
\usepackage[acronym,nomain]{glossaries}
\makeglossaries
\usepackage{nomencl}
\makenomenclature
\usepackage[title]{appendix}

\title{Low Complexity Radio Frequency Interference Mitigation for Radio Astronomy Using Large Antenna Array
}

\author{Zaid Bin Tariq$^{1}$, Teviet Creighton $^{2}$,  Louis P. Dartez$^{2}$, Naofal Al-Dhahir$^{1}$ and Murat Torlak $^{1}$ \\
$^{1}$Department of Electrical and Computer Engineering, University of Texas at Dallas, Richardson, TX 75080, USA\\
$^{2}$Center for Advanced Radio Astronomy, University of Texas, Rio Grande Valley, Brownsville, TX 78521, USA\\
zaidbin.tariq@utdallas.edu
}

\begin{document}
\maketitle

\begin{abstract}
With the ongoing growth in radio communications, there is an increased contamination of radio astronomical source data, which hinders the study of celestial radio sources. In many cases, fast mitigation of strong radio frequency interference (RFI) is valuable for studying short lived radio transients so that the astronomers can perform detailed observations of celestial radio sources. The standard method to manually excise contaminated blocks in time and frequency makes the removed data useless for radio astronomy analyses. This motivates the need for better radio frequency interference (RFI) mitigation techniques for array of size M antennas. Although many solutions for mitigating strong RFI improves the quality of the final celestial source signal, many standard approaches require all the eigenvalues of the spatial covariance matrix ($\textbf{R} \in \mathbb{C}^{M \times M}$) of the received signal, which has $O(M^3)$ computation complexity for removing RFI of size $d$ where $\textit{d} \ll M$. In this work, we investigate two approaches for RFI mitigation, 1) the computationally efficient Lanczos method based on the Quadratic Mean to Arithmetic Mean (QMAM) approach using information from previously-collected data under similar radio-sky-conditions, and 2) an approach using a celestial source as a reference for RFI mitigation. QMAM uses the Lanczos method for finding the Rayleigh-Ritz values of the covariance matrix $\textbf{R}$, thus, reducing the computational complexity of the overall approach to $O(\textit{d}M^2)$. Our numerical results, using data from the radio observatory Long Wavelength Array (LWA-1), demonstrate the effectiveness of both proposed approaches to remove strong RFI, with the QMAM-based approach still being computationally efficient.
\end{abstract}

% keywords can be removed

\section{Introduction}
Astronomers utilize radio telescopes to study celestial radio sources with time domain and transient astronomy gaining prominence for studying different phenomena. They utilize antenna arrays specially designed for receiving the radio waves from such sources. Due to high radio frequency interference (RFI), these observatories are built  at the quietest radio regions on Earth in order to avoid RFI from anthropogenic radio sources. These sources include but are not limited to satellites, television, broadcasting stations, radio, phone, and aircraft \cite{vafaei2020deep}. In most cases, time integration of the covariance matrices is not sufficient to reduce the effect of RFI and is constrained by the data handling capabilities of an observatory. In short, both imaging and studying the astronomical sources become difficult under heavy RFI. These become even more difficult for time domain or transient astronomy, where the astronomer needs to take quick measures to  trigger the detection of exotic transients  for detailed observation using different telescopes \cite{hessels2009radio}. In this case, the transient radio sky is least characterized on the seconds level timescale \cite{hessels2009radio}. Data from a large number of antennas helps with the sensitivity of these short transient events but requires large amount of computing resources \cite{hessels2009radio}. These computational limitations are due to various reasons, and faster RFI mitigation will provide one tool in a tool-box for astronomers to take a positive step towards the goal of studying short-lived radio events.

A standard approach to handle strong RFI challenge in radio astronomy is the excision of the time series corresponding to polluted RFI segments in the data recording \cite{wijnholds2014signal}. This excision requires the detection of the RFI-contaminated segments in the time series. There are three broad classes for this approach. The most common are the SumThreshold and cumsum approaches, where a segment is flagged if the data exceeds a threshold in the time-frequency domain \cite{vafaei2020deep},\cite{raza2002spatial},\cite{baan2004radio}. Next, there are the methods which include Singular Value Decomposition (SVD) or Principal Component Analysis (PCA) \cite{zhao2013windsat}, \cite{offringa2010post},\cite{AKERET201735}. Although transient RFI can be detected and the corresponding time series excised from the data, continuous RFI will make temporal excision of data useless. Depending upon the percentage of RFI contamination, we cannot simply detect and excise the RFI segment since long lasting RFI signals will require taking out most of recorded segment. Along with this being inefficient, short time recordings must be completely discarded which also deters studies that work with small time scales. Although there are frequency bands reserved for radio astronomy, some communication systems might not have adequate band pass filters to prevent signal leakage in these bands; and in any case, the study of radio astronomy should not be hindered by limited reserved bands \cite{sardarabadi2015spatial}. To prevent this loss of data, it is necessary to devise fast, efficient, and near real-time RFI mitigation techniques. Apart from allowing astronomers easy access to high-interference bands, these methods will allow the possibility of setting up the radio arrays closer to high-interference regions on Earth. Furthermore, with the expansion and advancement in communication technology, there is an increased demand for using the reserved frequency bands for communication purposes. Thus, there is a need for more sophisticated RFI mitigation methods to allow the coexistence and simultaneous utilization of the frequency bands for the purpose of both communication and  radio astronomy. 

Array signal processing methods are among the most prominent methods utilized for RFI mitigation. The idea is to decompose the received signal so that the size of the strong RFI subspace can be approximated and removed. For this purpose, the covariance between the signal from the $i^{th}$ and $j^{th}$ antenna in an array of antennas is used to form a covariance matrix $\textbf{R}$. Rank analysis of the covariance matrix from the radio telescopes is one of the proposed methods \cite{leshem2000multichannel}.  For the ideal noiseless case, the number of spatially distinct interferers is equal to the rank of the matrix. One can compare the eigenvalues of the covariance matrix to estimate the RFI subspace. However, in practice, the signal from the radio telescope contains noise, which is assumed to be below noise power level. In this case, we can compare the eigenvalues with the noise variance $\sigma ^{2}$ to find the RFI subspace. The noise covariance matrix is difficult to accurately approximate, and the RFI can likely be comparable to the background noise in terms of power, which might lead to inaccuracies \cite{leshem2000multichannel}. 

The minimum description length (MDL) principle is another approach for finding significant eigenvalues for the case when the noise variance is not available \cite{leshem2000multichannel}. It utilizes the geometric and arithmetic means of the eigenvalues of $\textbf{R}$ to estimate the number of interferers. MDL requires full eigenvalue decomposition, which increases the computational complexity of the overall RFI mitigation process. Furthermore the false alarm rate is unknown and not fixed under the assumption that the background signal apart from the RFI is not colored \cite{leshem2000multichannel}. Given that estimating an accurate noise covariance matrix is unlikely in this scenario, MDL might lead to inaccurate results, along with an increase in computational complexity. Spatial filtering is another approach to project out the RFI from $\textbf{R}$ \cite{raza2002spatial}, \cite{van2005performance}. The subspace spanned by the RFI is estimated using the eigenvalue and eigenvector pairs. Finding the size of this eigenspace is again dependent upon the rank analysis. The assumption are that the noise is white and that the visibility at a certain time scale is insignificant, otherwise the eigen decomposition (ED) is disturbed \cite{leshem2000multichannel}. In many cases, a clear distinction between the eigenvalues of the RFI subspace and the signal of interest (SOI) is very difficult, which leads to inaccuracies as stated earlier. Furthermore, the ED step has $O(M^{3})$ complexity, which might not be suitable for fast RFI mitigation for transient and time domain radio astronomy in monitor mode. In \cite{van2004signal}, the author proposes a Factor Analysis approach for detecting the number of interferers, which can then be utilized for estimating the projection matrix for filtering the RFI. The Factor Analysis Decomposition (FAD) approach can be seen as the generalization of ED, showing usefulness in cases where the noise covariance matrix, assumed to be diagonal, is unknown. An iterative alternating least-square approach is proposed for finding the FAD for the covariance matrix \cite{van2004signal}, which can also increase the time complexity of the overall RFI mitigation process .

In this work, we devise and investigate an approach that uses Fast-QMAM for faster mitigation of heavy RFI compared to the traditional eigenvalue approach thus adding a useful tool in the toolbox for astronomers. The method utilizes previously-collected data under similar radio-sky-conditions for determining the RFI subspace, which to the best of our knowledge has not been investigated previously. For example, the 1-day-separated celestial data at the same LST will have similar celestial radio-sky-conditions. Only the similar sky-conditions, i.e. similar declination and right ascension of radio sources, is required. These eigenvalue-based methods require all the eigenvalues of the covariance matrix. To find these eigenvalues, one needs to perform the ED of $\mathbf{R}$, which requirs $O(M^3)$ computations. Previously, a Krylov-subspace-based approach was proposed for computationally faster calibration of antennas \cite{sardarabadi2014application}. Our approach uses the Krylov-subspace-based Lanczos method for estimating the top $m$ eigenvalues using the Rayleigh-Ritz (RR) values. At each $m^{th}$ Lanczos step, our QMAM-based approach uses estimates of the top $m$ eigenvalues to calculate a statistic for detecting  the RFI subspace, leading to $O(dM^2)$ computation time. This reduction in computational complexity makes faster RFI mitigation possible, which, in turn, speeds up other processes, specially for time domain radio astronomy in monitor mode. Our numerical results, using the data from the radio observatory LWA-1, show that the proposed QMAM approach significantly improves the quality of the signal-of-interest while being computationally efficient in comparison to other similar methods that require computation of all the eigenvalues, such as MDL proposed in \cite{leshem2000multichannel}.

We make the following contributions: firstly, we investigate and adapt Geometric Mean to Arithmetic Mean (GMAM) ratio to QMAM allowing use of the estimates of the first $m$ eigen values for computationally faster RFI mitigation compared to the traditional eigenvalue approaches. In doing so, we show that previous LST data under similar sky-conditions can be utilized for removing RFI. Secondly, we show our results on real world LWA-1 observatory data sets consisting of 60$\,$ms recordings for imaging the radio-sky without combining the data over the time series. LWA-1 allows us to show the utility of the proposed solution on a unique telescope with 260 antenna stands operating the 10$\,$MHz to 88$\,$MHz frequency range. From the radio science point of view, this spectrum range is particularly interesting for radio astronomers attempting to deal with contaminated bands. Thirdly, we utilize a celestial-source-based Signal-to-Interference-plus-Noise Ratio (SINR) measure for evaluating the quality of achieved RFI mitigation. In doing so, we also propose an iterative SINR approach for RFI mitigation based on the availability of a celestial radio source to be used as a reference.

This paper is structured as follows: in Section \ref{sec:data_processing}, we describe the pre-processing of the data. In Section \ref{sec:methodology}, we present the two proposed RFI mitigation methodologies using array signal processing approaches i.e.1) fast QMAM algorithm for RFI mitigation, and 2) iterative-SINR approach utilizing celestial radio source as a reference. In Section \ref{sec:experiments}, we provide evaluation of our proposed methodology using LWA-1 data, and Section \ref{sec:conclusion} concludes the paper.

\section{Data Pre-Processing}

\label{sec:data_processing}
In this section, we present a general procedure for measuring and processing astronomy radio signals received by an $M$-element antenna array.  The signal received by each of the $M$ antenna elements at time $t$ can be represented by the vector

\begin{equation}
    \label{eq:Eigen_decompose_big}
\textbf{x}(t) = \left[\begin{array}{c}
  x_{1}(t) \\
  .\\
  .\\
  x_{\textup{M}}(t)\\
\end{array}\right].
\end{equation}

%\textit{\textbf{}}

Depending upon the array, $\textbf{{x}}(t)$ can be a wide-band or a narrow-band signal. The  signal from celestial sources is wide-band while RFI  is usually narrow-band.  Hence, it is appropriate to process the received signal for a specific narrow band to mitigate RFI. For this purpose, the output of the hardware generally is the empirical covariance matrix $\hat{\textbf{R}}_{f}$ for covariance of $\textbf{x}_f(t)$ where $f$ is  the central frequency of a particular narrow-band signal \cite{leshem2000multichannel} and is an estimate of the true covariance

\begin{equation}
    \textbf{R}_{f} = E\{\textbf{x}_{f}(t)\textbf{x}^{H}_{f}(t)\}.
\end{equation}

\iffalse
The covariance between the $i^{th}$ and $j^{th}$ antenna can be defined as the the convolution of $x_{i}(-t)$ and $x_{j}(t)$. Using the Fourier transform property, this convolution can be written  as follow:

 \begin{equation}
     x(-t) * x_{j}(t) = \overline{\mathcal{F}(x_{i}(-t))}\cdot \mathcal{F}(x_{j}(t))  \\
\end{equation}
\fi

For easy reference, we drop the frequency subscript for the rest of the paper. The geometry of the interferometer needs to be considered when performing the imaging of the celestial sources. The delays resulting from the geometry of antenna array, radio source, and cable length for each antenna stand are already captured in the signal from each stand. In order to compensate for this delay, we utilize the transformation in \cite{gaber2016navigating} to convert the local coordiantes in $(x,y,z)$ into $(u,v,w)$. These coordinates track the position of the celestial sources \cite{gaber2016navigating}. In this case, $\hat{w}$ is the vector towards the celestial source, $\hat{v}$ towards the celestial pole. The direction of $\hat{u}$ is defined such that $\hat{u} = \hat{v} \times \hat{w}$. 

Given the coordinates $(u,v,w)$, we can use them to compensate for the delays in the system. Specifically, because of the geometric positioning of the antennas in an array, a time varying delay is introduced to the signal between different antennas. There are mainly two types of delays in this case. The first delay is the geometric delay $\tau_{w}$.  This delay is incurred because of the placement of the antennas with respect to each other. Assuming a single celestial radio point source, the antenna placement changes the angle between an antenna and the radio source thus introducing delay between the signals. The second type of delay is observed due to the length of the cable between an antenna and the base station. In most arrays, the signal $\textbf{x}(t)$ from each antenna is transmitted to the central base station. Each antenna will have a different distance from the base station and hence a different cable length thus adding a delay of $\tau_{c}$. This delay can be calculated beforehand to be compensated at a later stage. 

Let $\tau_{c_{i}r}$ and $\tau_{g_{i}r}$ be the cable and geometric delays with respect to a reference antenna, respectively. The total delay $\tau_{i}$ for the $i^{th}$ antenna adds to the phase difference.  The phase difference with respect to an outrigger/reference antenna can be compensated through multiplying the factor $a_{i} = e^{-j2\pi f \tau_{i}}$ to the covariance values of the signal from the antennas under consideration. We can use the phase difference for each antenna to form a vector $\textbf{a} \in \mathbb{C}^{M}$ such that each $i^{th}$ element $a_{i}$ in $\textbf{a}$ is the phase between the $i^{th}$ and the reference antenna. It should be noted that the delay calculation and the corresponding steering vector $\textbf{a}$ is a function of the celestial coordinates (declination angle and right ascension), the longitude/latitude, and the Local Sidereal Time (LST) at the location of the antenna array. The beamfomring power for a particular declination $\delta_{i}$ and right ascension $\gamma_{i}$ is then computed using the beamforming operation $\textbf{a}^{H}\textbf{R}\textbf{a}$.

In this work, we utilize real world data from the Long Wavelength Array (LWA-1) to test our RFI mitigation method. We provide numerical results from our experiments for testing the QMAM approach. LWA-1 has a frequency range of 10-88$\,$MHz located in central New Mexico, USA. It consists of 260 cross dipole antennas. The outputs of the two orthogonal polarizations are separately digitized and enable all-sky imaging using beamforming. The data consists of the raw voltage outputs from each antenna. Each antenna stand generates two polarization data streams, i.e., X and Y polarizations. We use the data from the main array clustered together along with stand 258 as an outrigger. For each recording of LWA at a particular timestamp, we have 60$\,$ms of data at 15 minute intervals. The numerical results provided in this paper utilize this 60$\,$ms data at different LSTs unless otherwise stated. Working at this small time interval scale reduces the integration time of the covariance values, thus increasing the effect of the additive white noise. This also helps with testing the proposed approach given the small duration of data.

\section{RFI Mitigation Methodology}

\label{sec:methodology}

\subsection{System Model}
Given the measurement $\textbf{x}(t)$, we can formulate the measured signal model consisting of the RFI signal $\textbf{r}(t)$, the celestial radio source signal $\textbf{s}(t)$, and the noise/error signal $\textbf{n}(t)$. The combination of these signals makes up the measured signal $\textbf{x}(t)$ at the antennas and can be written as: 
\begin{equation}
\textbf{x}(t) = \textbf{r}(t)+ \textbf{s}(t)+\textbf{n}(t)    .
\end{equation}

For the purpose of radio astronomy, the goal is to remove or at least mitigate $\textbf{r}(t)$ from the measured signal while keeping the celestial source signal intact. The noise is assumed to be either comparable or smaller in terms of signal power and can thus be combined with the celestial signal for the purpose of our analysis, since our aim is to only mitigate the RFI. We can represent the data covariance matrix as:

\begin{equation}
    \textbf{R}= \textbf{R}_{r} + \textbf{R}_{s} = \textbf{B}\textbf{B}^{H} + \textbf{R}_{s}.
    \label{eq:data_covariance}
\end{equation}

In Eq. (\ref{eq:data_covariance}), $\textbf{B} \in \mathbb{C}^{M\times d}$ with $\text{rank}\{\textbf{B}\} =d$. Thus, we need to estimate  $\text{span}\{\textbf{B}\}$ to estimate the RFI subspace. Traditionally, this is possible using ED of the data  covariance matrix $\textbf{R}$, which can be represented as:

\begin{equation}
    \textbf{R} = \sum_{\textup{k}}\lambda_{k}\textbf{u}_{k}\textbf{u}_{k}^{\textup{H}} = \textbf{U}\Lambda \textbf{U}^{\textup{H}}.
     \label{eq:eigen_decomposition}
\end{equation}

In Eq. (\ref{eq:eigen_decomposition}), $\textbf{U}$ $\in$ $\mathbb{C}^{M \times M}$ is unitary such that $\textbf{U}^{\textup{H}}\textbf{U} = \textbf{I}$, with each column representing an eigenvector of $\textbf{R}$. 
The \textbf{$\Lambda$} $\in$ $\mathbb{R}^{M \times M}$ is such that the diagonal entries contain the eigenvalues in the descending order, i.e.,  $\lambda_{1}>\lambda_{2}>...>\lambda_{\textup{M}}$, with $\textbf{u}_{k}$ being the corresponding eigenvectors.  For the ideal RFI-free case, the ED in Eq. (\ref{eq:eigen_decomposition})  represents the celestial radio source signal and the background noise signal, which is assumed to be white. 

For the purpose of RFI mitigation, we can combine the background celestial radio signal and the noise as one signal of interest (SOI). By summing over the sequential covariance matrices, the effect of this noise can be minimized during imaging using beamforming. On the other hand, the effect of RFI cannot be mitigated in most cases by summing the covariance over time. Earth-based RFI sources have high signal power compared to the celestial radio sources. Under this assumption, we can further decompose the above ED for the covariance matrix as
 
\begin{align}
\label{eq:Eigen_decompose_big}
\textbf{R} &= \left[\begin{array}{cc}
  \textbf{U}_{d} & \textbf{U}_{s} \\
\end{array}\right]
\left[\begin{array}{cc}
  \Lambda_{d} & 0 \\
  0 & \Lambda_{s} \\
\end{array}\right]
\left[\begin{array}{cc}
  \textbf{U}_{d}^{\textup{H}} \\
  \textbf{U}_{s}^{\textup{H}}
\end{array}\right].
\end{align}

In the above decomposition, $\textbf{U}_{d}$ $\in$ $\mathbb{C}^{M \times d}$ and diagonal $\Lambda_{d}$ $\in$ $\mathbb{R}^{d \times d}$ are eigenvector and eigenvalue pairs for the interfering signal. Our goal is only to remove the RFI given by the number of interferers $d$ in the signal $\textbf{x}(t)$ and not to remove background sky sources. Hence, we have grouped the radio sky-source signal and the background noise signal such that it is represented by   $\textbf{U}_{s}$ $\in$ $\mathbb{C}^{M \times M-d}$ and diagonal $\Lambda_{s}$ $\in$ $\mathbb{R}^{M-d \times M-d}$. The above decomposition is far from the ideal case of  $\textbf{R} = \textbf{U}_{\textup{s}}\Lambda_{\textup{s}}\textbf{U}_{\textup{s}}^{H}$. However, we can cancel out the RFI  by approximating the RFI subspace and then subtracting it from the covariance matrix $\textbf{R}$ with the update
 \begin{equation}
      \textbf{R}_{d} = \textbf{R} -\textbf{u}_{1}\lambda_{1}\textbf{u}_{1}^{T} - \textbf{u}_{2}\lambda_{2}\textbf{u}_{2}^{T}…- \textbf{u}_{d}\lambda_{d}\textbf{u}_{d}
      .
      \label{eq:RFI_removal}
\end{equation}

Given that $\text{rank}\{\textbf{B}\} = d$, the RFI subspace is the $\text{rank}\{\textbf{B}\} = \text{rank}\{\textbf{U}_{\textup{q}}\}$ where $\textbf{U}_{\textup{q}} = [\textbf{u}_{1},...,\textbf{u}_{\textup{d}}]$ are the eigenvectors corresponding to the $d$ largest eigenvalues. It should be noted that the dimension $d$ of the RFI subspace is unknown and needs to estimated. In the following section, we show the utility of Geometric Mean to Arithmetic Mean (GMAM) metric for estimating $d$ given the availability of previous data under similar sky-conditions. Then, in order to improve the computational aspect of RFI subspace detection, we incorporate the Lanczos method and convert the GMAM ratio to a QMAM ratio for faster RFI mitigation.

\subsection{Geometric Mean to Arithmetic Mean }
\label{sec:GMAM}
For estimating the RFI subspace, one frequently used measure is GMAM given as:

\begin{equation}
L(\lambda_{\hat{d}+1},...,\lambda_{M})  = \frac{(\prod_{k=\hat{d}+1}^{M}\lambda_{k})^{1/(M-\hat{d})}}{\frac{1}{M-\hat{d}}\sum_{k=\hat{d}+1}^{M}\lambda_{k}}   .
\label{eq:GMAM}
\end{equation}

This measure is also utilized in different detection schemes like AIC \cite{bartlett1954note}, MDL \cite{1164557}, and the likelihood ratio test \cite{lawley1956tests}, \cite{xu1994fast}. The test statistic is based on the fact that the eigenvalues $\lambda_{k}$ for $d+1\leq k \leq M$ tend to cluster around the background noise $\sigma^{2}$ to test for $\hat{d} = d$ \cite{xu1991fast}. In general, GMAM will be approximately close to one, and a threshold needs to be established under the assumption of white noise. For the radio astronomy signal, the gap between the noise and the celestial source signal might even be more variable, and it is more suitable for the problem of RFI mitigation to treat the celestial signal along with system noise as colored noise. Thus, we can use similar celestial sky characteristics for RFI subspace detection. Given the weights $w_{k}$ such that $w_{d+1}\sigma^{2}>w_{d+2}\sigma^{2}>...>w_{d+s}\sigma^{2}=\sigma^{2}=...= \lambda_{M}$, where $\lambda_{k} =w_{k}\sigma^2 = w_{k}\lambda_{M}$   and $s$ is the size of the eigenvalues corresponding to the celestial source signal and $\sigma$ has multiplicity $M-d-s$, we can write the above GMAM ratio as:

\begin{equation}
L(\lambda_{\hat{d}+1},...,\lambda_{M})  = \frac{(\prod_{k=\hat{d}+1}^{M}\sigma^{2})^{1/(M-\hat{d})}(\prod_{k=\hat{d}+1}^{M}w_{k})^{1/(M-\hat{d})}}{\frac{\sigma^{2}}{M-d}\sum_{k=\hat{d}+1}^{M}w_{k}}   
\end{equation}

\begin{equation}
L(\lambda_{\hat{d}+1},...,\lambda_{M})  = \frac{C(\prod_{k=\hat{d}+1}^{M}\sigma^{2})^{1/(M-\hat{d})}}{\frac{\sigma^{2}}{M-d}}   
\label{eq:GMAM_simplified_1}
\end{equation}

where $C = \frac{(\prod_{k=\hat{d}+1}^{M}w_{k})^{1/(M-\hat{d})}}{\sum_{k=\hat{d}+1}^{M}w_{k}}$. Eq. (\ref{eq:GMAM_simplified_1}) can then be simplified as:

\[L(\lambda_{\hat{d}+1},...,\lambda_{M}) = (M-\hat{d})C.\]

The hypothesis of this work is that the ratio $C$ from the previously available radio astronomy data remains the same as the data under consideration for RFI mitigation, given similar sky-conditions (same LST but different day). In the following sections, we show that the cleaned data from a previous LST with similar sky-conditions can be utilized for computing an estimate of the weights. It should be noted that even if we have the weights, the measure in Eq. (\ref{eq:GMAM_simplified_1}) will still require all the eigenvalues using computationally expensive ED. This brings us to our second hurdle of making this measure computationally less costly. In the following section, we utilize the above form and reasoning of the GMAM ratio to find the Quadratic Mean to Arithmetic Mean (QMAM) measure  along with the Lanczos method to reduce the computational complexity to $\mathbb{O}(dM^{2})$.

\subsection{Lanczos Method }

The Lanczos method provides an estimate of the top $m$ eigenvalues and eigenvectors of the covariance matrix of size $M$. Assuming that the dimensions of the RFI subspace $d \ll M$, we can  utilize the first $m$ eigenvalues to determine $d$ by using a test statistic. In this work, we present two methods for finding $d$ using the Lanczos algorithm with overall complexity of $\mathbb{O}(d^{2}M)$. In this section, we briefly describe the use of Krylov subpaces and the Lanczos method to find the Rayleigh-Ritz values that estimate the $\text{span}\{\textbf{U}_{\textup{q}}\}$, i.e., the RFI subspace.

Assuming that we can find  an initial $M \times 1$ starting vector $\textbf{f}$ and given the covariance matrix $\textbf{R}$, the Krylov matrix is represented by $K^{m}(\textbf{R},\textbf{f}) = [\textbf{f},\textbf{R}\textbf{f},...,\textbf{R}^{m-1}\textbf{f}]$ where $K^{m}(\textbf{R},\textbf{f}) \in \mathbb{C}^{M \times m}$. This matrix spans the Krylov subspace, i.e. $\mathcal{K}^{m}(\textbf{R},\textbf{f}) = \text{span}\{K^{m}(\textbf{R},\textbf{f})\}$. Given the Krylov subspace, it can be shown that for a Hermitian matrix \textbf{R}, $\mathcal{K}^{m}(\textbf{R},\textbf{f}) ={K}^{m}(\textbf{R}-\rho \textbf{I},\textbf{f}) = {K}^{m}(\textbf{B}\textbf{B}^{\textup{H}},\textbf{f}) $, where $\rho$ is any scalar \cite{xu1991fast}. Furthermore, it can be shown that for $\textbf{R}$ having $d < M-1$ distinct eigenvalues with $M-d$ repeated eigenvalues, then we have the following mathematical results \cite{xu1991fast}

\begin{enumerate}

    \item ${K}^{d+1}(\textbf{R},\textbf{f}) = span\{\textbf{U}_{\textup{q}},\textbf{f}\} = span\{\textbf{B},\textbf{f}\}$
    \item ${K}^{m}(\textbf{R},\textbf{f}) = {K}^{d+1}(\textbf{R},\textbf{f}), m \geq d+1$.
\end{enumerate}

For the detailed proof of these results, we refer the reader to the work in \cite{xu1991fast}. These results show that $\mathbf{R}$ acts like a $\text{rank}\{d\}$ matrix in its Krylov subspace and that the Krylov subspaces stops growing after $d+1$ steps of finding the Krylov subspace \cite{xu1991fast}. Hence, this fact can be used to find the RFI subspace of size $d$ without knowing all the eigenvalues, thus helping in reducing the overall complexity for finding the signal subspace. 

However, finding Krylov basis vectors requires $\mathbb{O}(d^{2}M)$ flops to re-orthogonalize all the previous basis vectors at each step. This particular step for finding the Krylov subspaces increases the overall complexity of finding the RFI subspace. The Lanczos method mitigates this issue and helps to find an orthonormal basis in $\mathbb{O}(dM)$ \cite{xu1991fast}. To find the orthonormal basis $\textbf{P}_{j} = [\textbf{p}_{1},...,\textbf{p}_{j}]$ for $\mathcal{K}^{j}(\textbf{R},\textbf{f})$, Lanczos showed that we only need to orthogonalize the vector $\textbf{R}\textbf{p}_{j}$ against $\textbf{p}_{j-1}$ and $\textbf{p}_{j}$, i.e., $\textbf{p}^{\textup{H}}_{i}\textbf{R}\textbf{p}_{i} = 0, i<j-1$ \cite{xu1991fast}. The algorithm used to find the estimated RR values ($\theta^{m}_{k}$) and vectors ($\textbf{y}^{m}_{k}$) is summarized in Algorithm \ref{algo:Lanczos}. At each $m^{th}$ step, we can find a tridiagonal $\textbf{T}_{m} = \textbf{P}^\textup{H}_{m}\textbf{R}\textbf{P}_{m}$ with the matrix structure shown in (\ref{eq:T_m}).  The RR value $\theta^{m}_{k}$ is the eigenvalue of the matrix $\textbf{T}_{m}$. In the following subsections, we describe two approaches, Fast Subspace Decomposition (FSD) and Source-based RFI Mitigation (SRM), for low-complexity RFI removal using the RR values and vectors.

\begin{align}
\label{eq:T_m}
\textbf{T}_{m} &= 
\left[\begin{array}{ccccc}
  \alpha_{1} & \beta_{1} &  & &\\
  \beta_{1} & \alpha_{2}& \beta_{2} & & \\
   & \beta_{2}& . & .& \\
   & & . &\alpha_{m-1} & \beta_{m-1}\\
   & &  &\beta_{m-1} &\alpha_{m} \\
\end{array}\right]
% \left[\begin{array}{cc}
%   U_{q}^{H} \\
%   U_{s}^{H}
% \end{array}\right]
\end{align}

\begin{algorithm}[t]

$\bold{Input}$ $\textbf{r}_{0}=\textbf{f}, \beta_{0} = ||\textbf{r}_{0}||$, For $j=1,...,m$
\begin{enumerate}
    \item $\textbf{p}_{j}:\textbf{p}_{j} \xleftarrow{}\textbf{r}_{j-1}/\beta_{j-1}$ Normalize $\textbf{r}_{j-1}$
    \item $\textbf{v}_{j}:\textbf{v}_{j} \xleftarrow{} \textbf{R}\textbf{p}_{j}$
    \item$\textbf{r}_{j}\leftarrow{} \textbf{v}_{j}-\textbf{p}_{j-1}\beta_{j-1}(\textbf{p}_{0}= \boldsymbol{0})$ (Orthogonalize $\textbf{v}_{j}$)

    \item $\alpha_{j}:\alpha_{j}\leftarrow{} \textbf{p}^{\textup{H}}_{j}\textbf{r}_{j}$

    \item $\textbf{r}_{j}:\textbf{r}_{j} \leftarrow{}\textbf{r}_{j}-\textbf{p}_{j}\alpha_{j}$ (Orthogonalize $\textbf{r}_{j}$ with respect to $\textbf{p}_{j}$)

    \item $\beta_{j}:\beta_{j}\leftarrow{} ||\textbf{r}_{j}||$ (Norm of $\textbf{r}_{j}$)
    
\end{enumerate}

\caption{Lanczos Algorithm}
\label{algo:Lanczos}
\end{algorithm}

It should be noted that there will still be an error in the estimated $\text{span}\{B\}$ given that we use the sample covariance matrix rather than the ideal covariance matrix. The error can be represented as follows \cite{xu1994fast}:

\begin{equation}
\text{span}\{\textbf{B}\} = \text{span}\{\textbf{e}_{k}\}^{d}_{k=1}   = \text{span}\{\hat{\textbf{e}}\}^{d}_{k=1} +\mathbb{O}(N^{1/2})
\end{equation}

The error is dependent upon the number of samples $N$ used for estimating the covariance matrix $\textbf{R}$. Taking this observation under consideration, the RR values and vectors are still able to estimate the eigenvalues and vectors of a sample covariance matrix $\hat{\textbf{R}}$ with in $\mathbb{O}(N^{-(m-d)})$ and $\mathbb{O}(N^{-(m-d)/2})$, respectively \cite{xu1994fast}. In the next section, we use the resulting RR values generated using the Lanczos method to estimate the size of RFI subspace to subtract out the RFI from the signal.

\subsection{Fast QMAM-based RFI Mitigation}

As previously mentioned, RFI removal methods like MDL or eigenvalue testing, utilize all the eigenvalues of the covariance matrix for estimating the size of the RFI subspace. The requirement for all the eigenvalues of the covariance matrix $\textbf{R}$ further increases the computational complexity of removing RFI from the radio signal $\textbf{x}(t)$ to $\mathbb{O}(M^3)$. As mentioned earlier, the Lanczos methods and its variants allow us to estimate the top $m$ eigenvalues and eigenvectors of the covariance matrix in $\mathbb{O}(mM^2)$. Given that we can utilize the top $m$ eigenvalues for finding the RFI subspace $\textbf{U}_{d}$, we can reduce the computational complexity of RFI removal to $\mathbb{O}(dM^2)$ where $d$ is the actual size of RFI subspace. 

For the case of the geometric mean in GMAM in Eq. (\ref{eq:GMAM}), we can not use the determinant of $\hat{\textbf{R}}$ since the determinant of an $M \times M$ matrix has $\mathbb{O}(M^3)$ complexity. We can replace the numerator of the GMAM ratio into a quadratic mean as

\[    \sqrt{\frac{1}{M-\hat{d}}\sum_{k = d+1}^{M}\lambda_{k}^{2}}.\]

The motivation to utilize the ratio QMAM comes from fast subspace decomposition algorithm where QMAM is given by:

\begin{equation}
    QMAM = \eta(\lambda_{d+1},...,\lambda_{M}) = \frac{\sqrt{\frac{1}{M-\hat{d}}\sum_{k = d+1}^{M}\lambda_{k}^{2}}}{\frac{1}{M-\hat{d}}\sum_{k=\hat{d}+1}^{M}\lambda_{k}}. 
    \label{eq:QMAM_ratio}
\end{equation}

The aim is to convert the above ratio to write the resulting measure in terms of the RR values, which does not require computationally expensive ED \cite{xu1994fast}. Thus, we can begin by replacing the arithmetic sum as:

\begin{equation}
    \sum_{k = \hat{d}+1}^{M} \hat{\lambda}_{k}= Tr(\hat{\textbf{R}}) - \sum_{k = 1}^{d} \hat{\lambda}_{k}
\end{equation}

\noindent
where  $\lambda_{k}$ can be replaced by $\hat{\lambda}_{k}  = \theta_{k} + \mathbb{O}(N^{-(m-d)})$ \cite{xu1994fast}. This in turn helps us to compute the arithmetic mean without finding all the eigenvalues, which requires ED. Now utilizing the identity $||\hat{\textbf{R}}||_{F}^{2} = \sum_{k=1}^{M}\lambda_{k}^2$  ($||\cdot||_{F}$ = denotes the Forbenius norm) and replacing $\lambda_{k}$ by the RR value $ \theta_{k}$, we have the following form for the quadratic sum of the eigenvalues, which is also computationally less costly: $\sum_{k = d+1}^{M}\lambda_{k}^{2} = ||\hat{\textbf{R}}||_{F}^{2} - \sum_{k=1}^{d}\theta_{k}^{(m)^{2}}+ \mathbb{O}(N)^{-(m-d)}$. Thus we can finally write the QMAM ratio in terms of the RR values as follows:

\begin{equation}
    \eta(\lambda_{d+1},...,\lambda_{M}) \approx \eta(\theta_{1}^{(m)},...,\theta_{\hat{d}}^{(m)}) = \frac{\sqrt{\frac{1}{M-\hat{d}}(||\hat{\textbf{R}}||_{F}^{2}-\sum^{\hat{d}}_{k=1}\theta^{(m)^{2}}_{k})}}{\frac{1}{M-\hat{d}}(Tr(\hat{\textbf{R}})-\sum^{\hat{d}}_{k=1}\theta^{(m)^{2}}_{k})}.
    \label{eq:QMAM_ritz_values}
\end{equation}

Note that Eq. (\ref{eq:QMAM_ratio}) and (\ref{eq:QMAM_ritz_values}) are the same. As pointed out in \cite{xu1994fast}, under the assumption of white noise $\sigma^2 \textbf{I}$ and the remaining eigenvalue multiplicity of $M-d$, the ratio $\eta(\lambda_{d+1},...,\lambda_{M}) = 1$. But because of the celestial signal, the multiplicity of the remaining eigenvalues is no longer approximately $M-d$ and there is an added factor such that $\eta_{ideal}(\lambda_{d+1},...,\lambda_{M}) = \epsilon \times \eta(\lambda_{d+1},...,\lambda_{M})$ where

\begin{equation}
     \frac{1}{\epsilon}=  \frac{\sqrt{\frac{1}{M-d}\sum_{k = d+1}^{M}w_{k}^{2}}}{\frac{1}{M-d}\sum_{k=d+1}^{M}w_k} .
    \label{eq:QMAM_ratio_weights}
\end{equation}

Given a good approximation of $\epsilon$, we are able to convert QMAM to the ideal scenario, which helps with finding $d$. Eq. (\ref{eq:QMAM_ratio_weights}) helps us to see the comparison of radio-sky with similar conditions. The assumption is that the distribution of $w_{k}$ for $d\leq k \leq M$ and hence the ratio in (\ref{eq:QMAM_ratio_weights}) is going to be similar for similar sky-conditions and system noise power. Hence, with the already computed threshold $\epsilon$ for previous data with same sky-conditions (as mentioned earlier, similar sky-conditions are usually found at similar declination and right ascension of the source), we can define the threshold $\phi_{\hat{d}}$ such that:

\begin{equation}
    \phi_{\hat{d}} = log(\epsilon \times \eta(\theta_{1}^{(m)},...,\theta_{\hat{d}}^{(m)}))
    \label{eq:statistic_QMAM}
\end{equation}

\begin{figure*}
\centering
\begin{subfigure}{0.33\textwidth}
  \includegraphics[width=\linewidth]{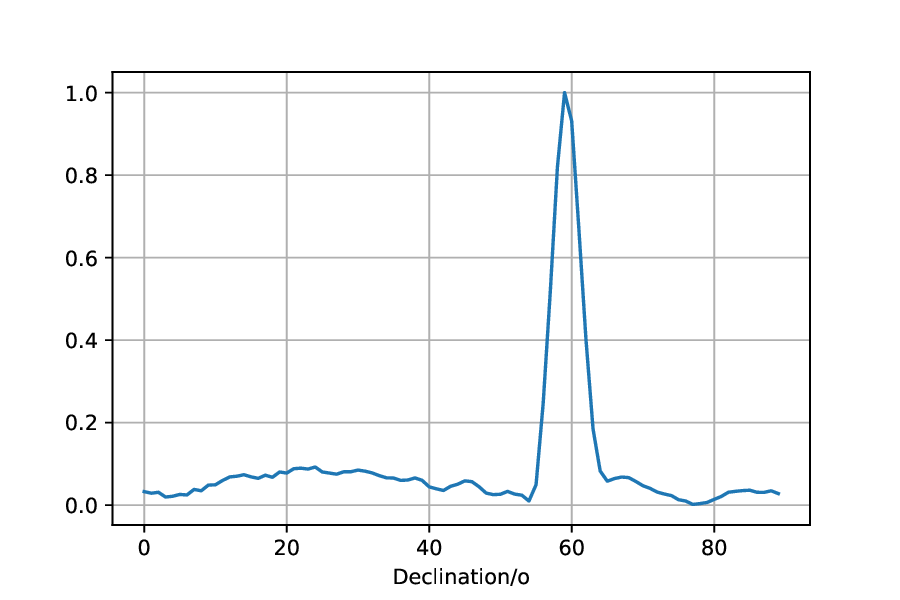}
  \caption{RFI-free band (41MHz)}
  \label{fig:sub1}
\end{subfigure}%
\begin{subfigure}{0.33\textwidth}
  \includegraphics[width=\linewidth]{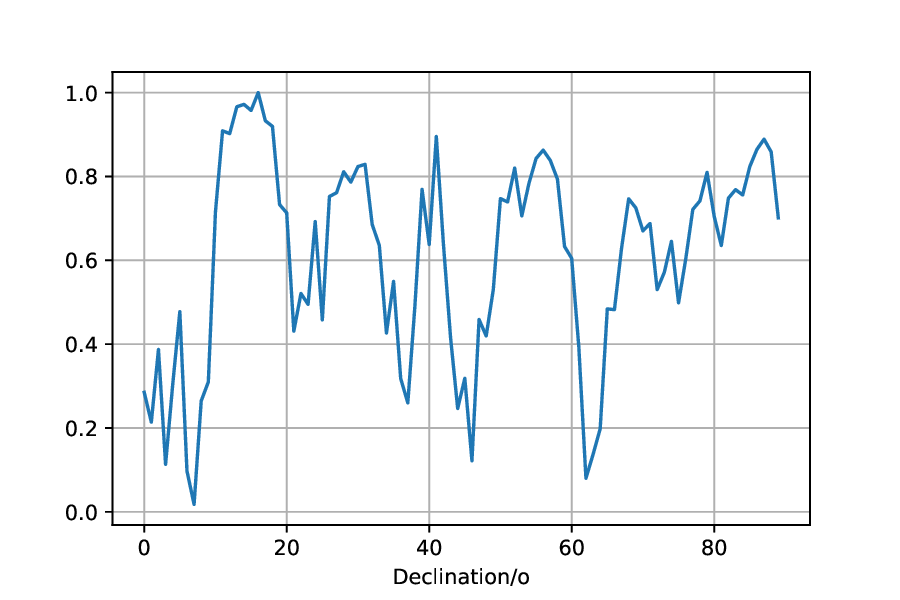}
  \caption{Before RFI mitigation (27MHz)}
  \label{fig:sub3}
\end{subfigure}
\begin{subfigure}{0.33\textwidth}
  \includegraphics[width=\linewidth]{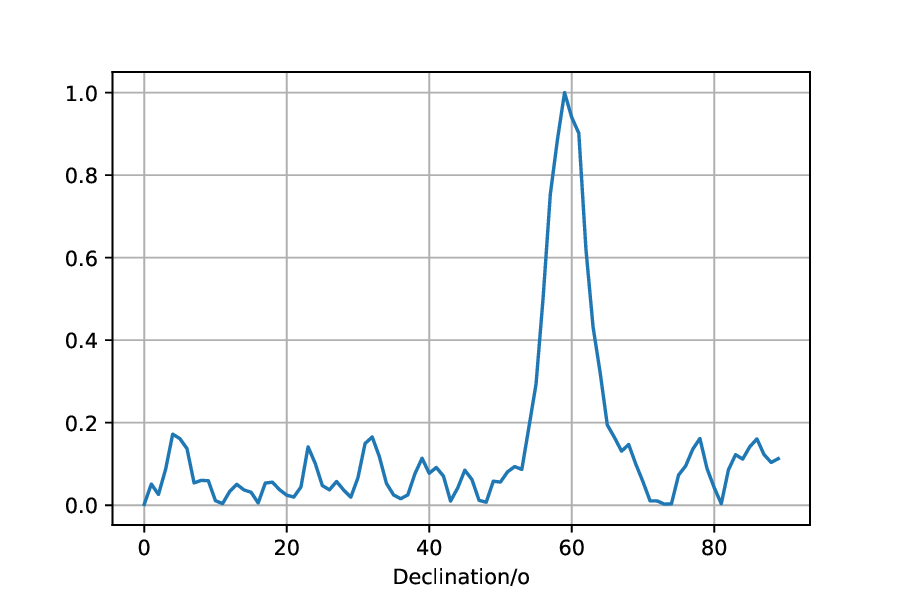}
  \caption{After RFI mitigation (27 MHz)}
  \label{fig:sub2}
\end{subfigure}

\caption{The normalized beamforming output vs. declination where the right ascension is set to $23^{h}23^{m}$ (Cassiopeia-A as the source). We can see that the larger beam width is narrower for higher frequencies (41$\,$MHz) compared to lower frequencies (wider span at 27$\,$MHz). The plot also shows the comparison of the resulting drift scan before (\ref{fig:sub3}) and after (\ref{fig:sub2}) RFI mitigation using the proposed methodology in this paper. }
\label{fig_beamforming_example}
\end{figure*}

\begin{algorithm}[t]

% \KwIn{$\gamma$}
\begin{enumerate}
    \item  Assume $ \hat{d} = 1$
    \item Set the hypothesis $H_{o}: d = \hat{d}$.
    \item Calculate the test static $\phi_{\hat{d}}$ in (\ref{eq:statistic_QMAM}).
    \item if $\phi_{\hat{d}} < 0$ accept $H_{o}$. Otherwise $\hat{d} = \hat{d}+1$. Given $\hat{d}>m-d$ continue with $m+1$ Lanczos step.
\end{enumerate}
\caption{Detection Using QMAM}
\label{algo:QMAM_detection_algo}
\end{algorithm}

Thus, we use $\phi_{\hat{d}}$ at the Lanczos $m^{th}$ step for finding the dimensions $d$ of the RFI subspace as shown in Algorithm \ref{algo:QMAM_detection_algo}.  We start by setting up a hypothesis test that $H_{o}:d = \hat{d}$. Then, we keep on evaluating $\phi_{\hat{d}}$, each time checking if $\phi_{\hat{d}} < 0$ to accept the hypothesis $H_{o}$. It should be noted that the previous data from multiple LST with the same sky-condition can be used for increasing the reliability of the factor $\epsilon$. This data can be obtained from previous days under similar sky-conditions i.e similar source locations with respect the array. Given the data from LST under similar sky-conditions $t_{1},...,t_{K}$, we can also estimate the corresponding $\epsilon$ as $\epsilon_{t_{1}},...,\epsilon_{t_{K}}$ and combine them to get the final estimate of $\epsilon_{f} = \frac{1}{K}\sum{\epsilon_{t_{k}}}$. We call this approach combined-QMAM, and show the merits of the technique in the following section.

The overall RFI removal process can be completed using the following three steps:
\begin{enumerate}
    \item Carry out the $m^{th} $ Lanczos step to compute the RR values $\theta^{m}_{k}, k = 1,...,m$.
    \item Use Algorithm \ref{algo:QMAM_detection_algo} to check $H_{o}$. If $H_{0}$ is true, then go to Step 3. Otherwise, continue with $m:=m+1$ Lanczos step and go to Step 1.
    \item Compute the estimate $\textbf{y}_{k}$ of the top $d$ eigenvectors corresponding to $\mathbb{K}^{m}(\hat{\textbf{R}},\textbf{f})$ to estimate the RFI subspace $\hat{\textbf{U}}_{q} = [\textbf{y}^{m}_{1},...,\textbf{y}^{m}_{d}]$. Cleaned covariance matrix $\textbf{R}_{d} = \hat{\textbf{R}} - \hat{\textbf{U}}_{q}\Theta \hat{\textbf{U}}^{\textup{H}}_{q} $ where $\Theta$ is the diagonal matrix containing the RR values in descending order.
    
\end{enumerate}

It should be noted that the estimated factor $\epsilon$ is calculated under the assumption that the previous data under similar sky-conditions is either RFI free or cleaned from RFI using an existing RFI mitigation technique. We can use recording which has very low RFI percentage contamination allowing to excise the time series segments to act as a template. For this work, we utilize the iterative-SINR approach presented in the next subsection.

\subsection{Iterative Source to Interference Noise Plus Ratio (Iterative-SINR)}
 \begin{figure*}[]
% \begin{minipage}{\textwidth}
\begin{center}

\begin{tikzpicture}
  \node (img)  {\includegraphics[width=0.5\linewidth]{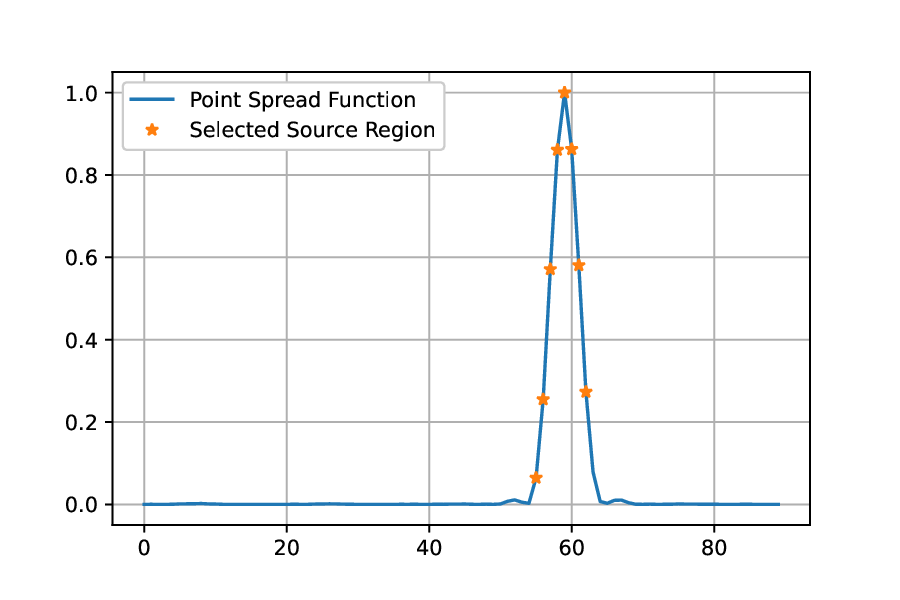}};
  \node[below=of img, node distance=0cm,xshift = 0.5cm, yshift=1.5cm,font=\color{black}] {Declination/$^\circ$};
%   \node[left=of img, node distance=0cm, rotate=90, anchor=center,yshift=-1cm,font=\color{black}] {SINR/dB};
 \end{tikzpicture}
\end{center}%
\caption{Plot of the normalized ambiguity function (i.e., point spread function) against declination for the 41$\,$MHz band. The signal corresponds to the region marked by the star which is selected once the magnitude of the function crosses above and below the mean of the point spread function.}
\label{fig_beamforming_example_ambiguity}
\end{figure*}

\begin{figure}[]
\centering
\includegraphics[width=0.6\linewidth]{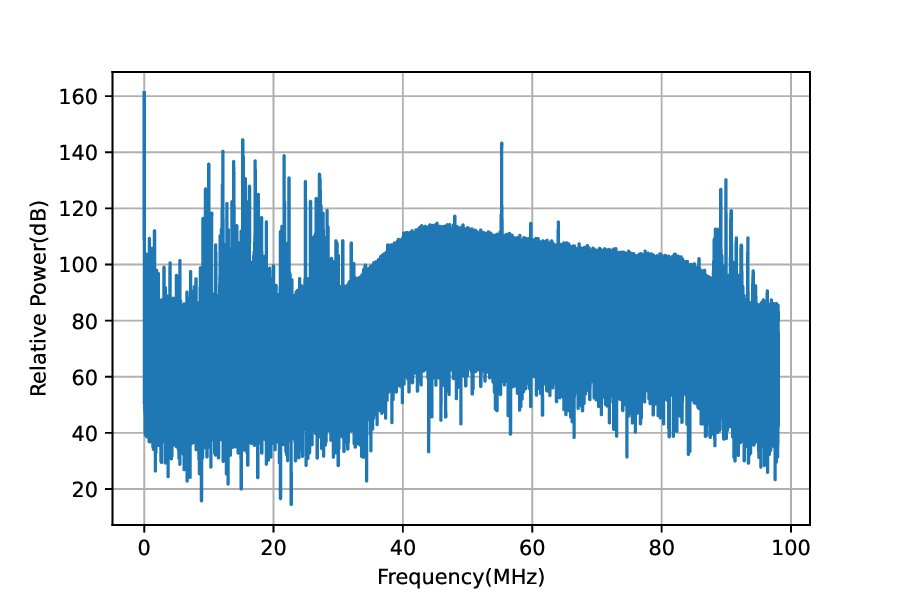}
\caption{The spectrum of the signal from Stand 258 at LWA array at $0^{h}14^{m}$ LST. LWA captures 10-88 MHz signals in the test mode. }
\label{fig:spectrum}
\end{figure}

In the previous subsection, we presented the QMAM-based RFI mitigation algorithm that uses the previous data under similar sky-conditions. In this subsection, we present a method that utilizes a strong celestial radio point source as a reference for RFI mitigation. An important factor to evaluate the RFI mitigation is visual reinforcement and in order to quantify the quality of RFI removal, we use the  SINR metric

\begin{equation}
    SINR =  \frac{\frac{1}{L_{\delta_{s}}} \sum\limits_{\delta \in \delta_{s}} \textbf{a}_{\delta}^{\textup{H}}\textbf{R}_{d}\textbf{a}_{\delta} - Tr(\textbf{R}_{d})}{\frac{1}{L_{\delta}}  \sum\limits_{\delta_i = \delta_{0}}^{\delta_{L}}\textbf{a}_{\delta_{i}}^{H}\textbf{R}_{d}\textbf{a}_{\delta_{i}}-Tr(\textbf{R}_{d})}.
\label{eq:SNR}
\end{equation}

In the above equation, $\textbf{a}_{\delta}$ is the vector with phase difference for each antennas in the array toward declination $\delta_{i}$, given that we are focusing at a source at a certain right ascension as described in section \ref{sec:data_processing}, with $L_{\delta}$ being the total length of set $\delta = [\delta_{0},...,\delta_{L}]$. Since we are looking at the strongest radio source and mitigating the stronger RFI, computing SINR for one right ascension is sufficient in this case which corresponds to the strongest celestial radio source. Also note that the SINR measure in Eq (\ref{eq:SNR}) is similar to the conventional covariance-based SINR with SOI being the selected celestial source and interferers being along the declination . In this SINR metric, we are taking the ratio of the power of the source to that of the average background for a pre-defined declination range in which the strongest source at a given LST recording can be found. For example, the strongest source at LST $23^h23^m$ is Cassiopeia-A (Cas-A) and the ratio of the magnitude corresponding to the declination of Cas-A ($58.8^{\circ}$) is used. However, it should be noted that a radio sky-source, which is used as a source in Eq. (\ref{eq:SNR}), is also radiating energy around its declination at $\delta_{s}$ due to finite spatial discrimination (beam width) of the array. This can be explained with the help of Figure \ref{fig_beamforming_example} which shows the beamforming output with the beam directed at Cas-A ($\delta_{s} = 58.8^{\circ}$). Although $\delta_{s} = 58.8^{\circ}$ is the declination of Cas-A, the beamforming result will still be picking up the source signal given that it is within a certain deviation of the source declination. This is also observable from Figure \ref{fig:sub1}, where the peak attenuates to the background power level within a few degrees of the source declination. This is the case with most of the radio sky-sources depending upon the frequency band of the observed signal $\textbf{x}(t)$. The beam width becomes wider for lower frequency bands as seen from Figure \ref{fig:sub2}. Hence, it is necessary for us to measure the span of declination of the source signal $\delta_{s}$ to be considered in setting up the SINR measure for testing the quality of RFI removal.

In order to define this range $\delta_{s}$, we define the point spread function in terms of beamforming as $\textbf{b}_{\delta} = [b_{\delta_{0}},...,b_{\delta_{L}}]$ where $b_{\delta_{l}} = \textbf{a}^{H}_{\delta_{l}}\textbf{R}_{s}\textbf{a}_{\delta_{l}}$ represents the magnitude of the point spread function at each declination and
$\textbf{R}_{s}= \textbf{a}_{s}\textbf{a}_{s}^{H}$ is imitating the covariance of an ideal source such that the phase outputs are perfectly aligned. Figure  \ref{fig_beamforming_example_ambiguity} shows the beamforming output using the point spread function for 41$\,$MHz with Cas-A as the beamforming source. It can be observed that the declination span of the peak represents the span of the ideal source, and this can be utilized for finding the declination range $\delta_{s}$ for Eq. (\ref{eq:SNR}). To identify source points from $b_{\delta}$, we can define $\textbf{h}_{\delta} = [h_{\delta_{0}},...,h_{\delta_{L}}]$ such that:

\begin{equation}
    h_{\delta_{l}} = \frac{1}{b_{\delta_{l}}-\Bar{b_{\delta}}}\text{max}(0,b_{\delta_{l}}-\Bar{b_{\delta}})
\end{equation}

In the above equation, $\bar{b_{\delta}}$ is the mean of the point spread function defined over the declination. Thus, the value of $h_{\delta_{l}}$ will be equal to $1$ if the point spread function  at a particular declination $\delta_{l}$ is greater than the mean of the overall point spread function across the considered range of declination. The declination range for which $h_{\delta{l}}=1$ is taken to be the source declination range $\delta_{s}$. Given the availability of suitable celestial source, we can iteratively solve the following problem to get the size of RFI subspace $d$:

 \begin{equation}
\centering
\begin{aligned}
\quad \underset{\hat{d}}{\text{max.}} & \quad  SINR(\textbf{R}_{\hat{d}},\delta)\\ 
\text{s.t.}  & \quad 0 \leq \hat{d} \leq M,   \ \text{and} \ \delta_{o} \leq \delta \leq \delta_{L-1}.
\end{aligned}
\label{eq_problem_general}
\end{equation}

Given $d$, we can use Eq. (\ref{eq:RFI_removal}) to calculate the RFI-mitigated covariance matrix given that we already have the eigenvalues and vectors using ED. Although this method is computationally more expensive compared to the QMAM approach, given the availability of a celestial radio source, this method allows us to find the best possible setting with the SINR measure. In the next section, we present results for the SINR and QMAM algorithms and compare them with the existing MDL approach for finding significant eigenvalues for RFI mitigation.

\section{Experiments and Evaluation}
\label{sec:experiments}

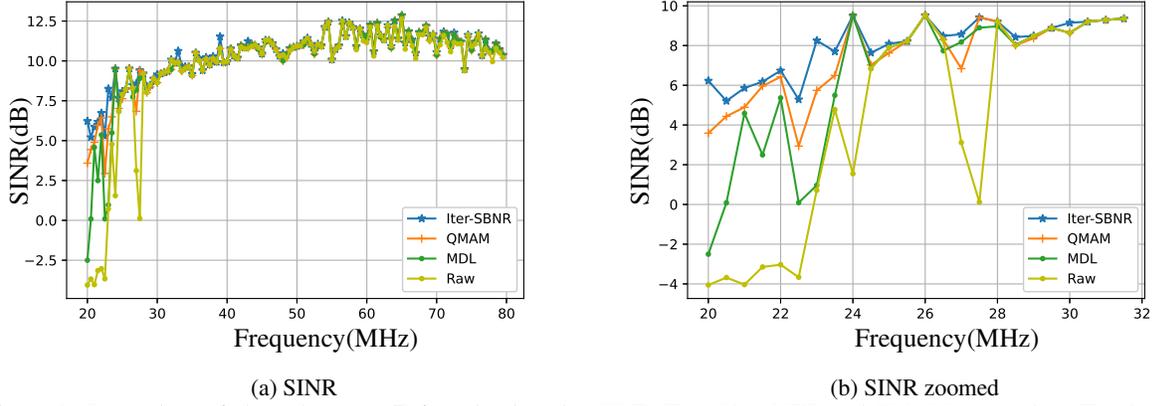
\begin{figure*}[t]
\begin{minipage}{0.5\textwidth}
\begin{tikzpicture}
  \node (img)  {\includegraphics[width=0.95\linewidth]{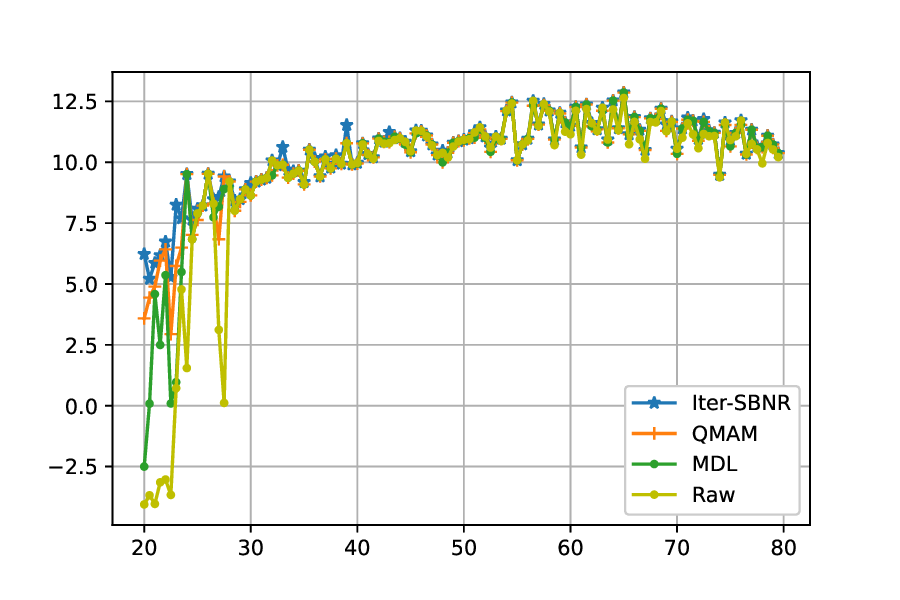}};
  \node[below=of img, node distance=0cm, xshift = 0.5cm,yshift=1.5cm,font=\color{black}] {Frequency(MHz)};
  \node[left=of img, node distance=0cm, rotate=90, anchor=center,yshift=-1.5cm,font=\color{black}] {SINR(dB)};
 \end{tikzpicture} 
  \subcaption{SINR}
\label{fig:5all}
\end{minipage}%
\begin{minipage}{0.5\textwidth}
\begin{tikzpicture}
  \node (img)  {\includegraphics[width=0.95\linewidth]{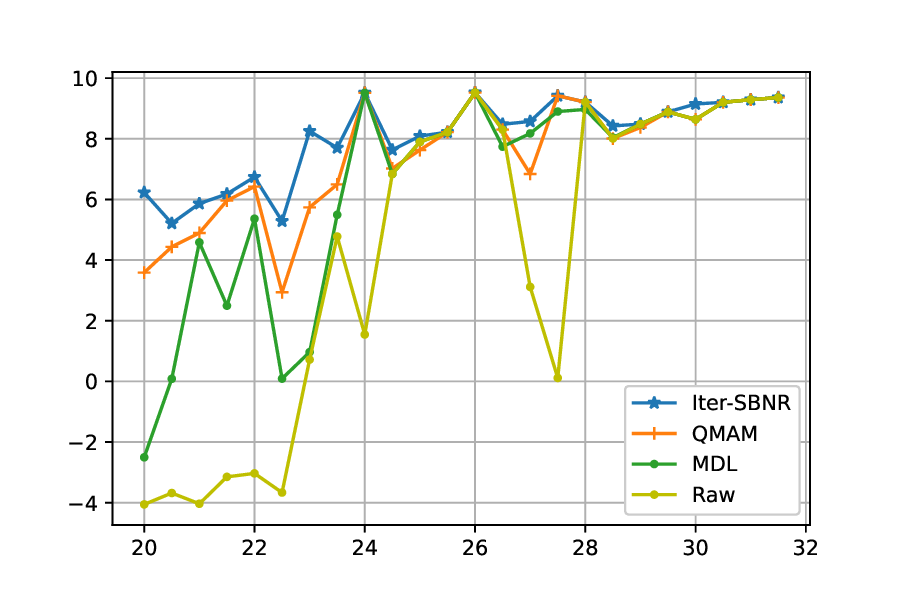}};
  \node[below=of img, node distance=0cm,xshift = 0.5cm, yshift=1.5cm,font=\color{black}] {Frequency(MHz)};
  \node[left=of img, node distance=0cm, rotate=90, anchor=center,yshift=-1.5cm,font=\color{black}] {SINR(dB)};
\end{tikzpicture}
 \subcaption{SINR zoomed}
  \label{fig:5zommed}

\end{minipage}%
\caption{Comparison of cleaned vs. raw $\textbf{R}$ for using iterative SINR (Eq. (\ref{eq_problem_general})), MDL and  QMAM approaches. For the case of QMAM, the data used for estimating $w_{k}$ and $\epsilon$ is collected at $5h11m$ LST with RFI being removed from data 1 day apart at $5h0m$ LST.}
\label{fig:1day}
\end{figure*}

\begin{figure*}[hbt]
\begin{minipage}{0.5\textwidth}
\begin{tikzpicture}
  \node (img)  {\includegraphics[width=0.95\linewidth]{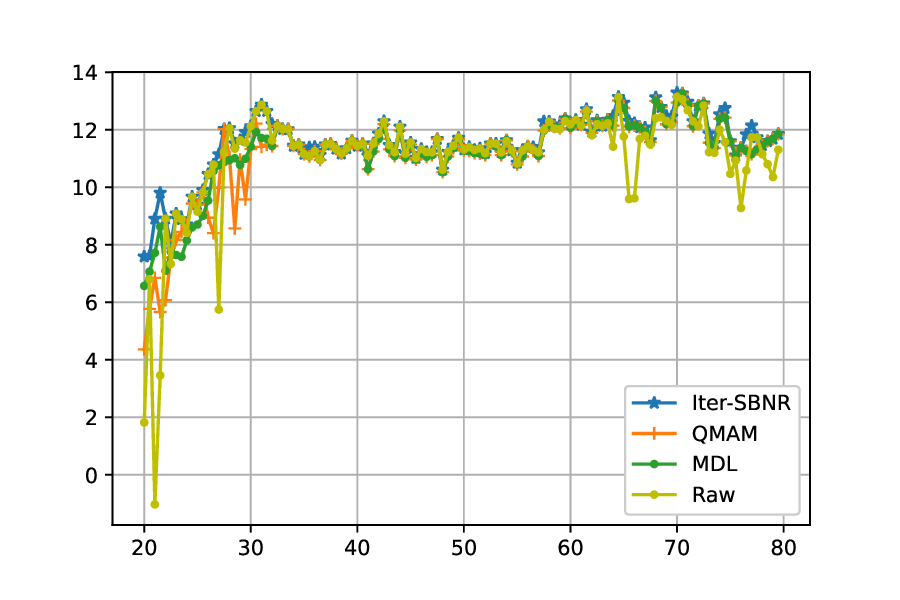}};
  \node[below=of img, node distance=0cm, xshift = 0.5cm,yshift=1.5cm,font=\color{black}] {Frequency(MHz)};
  \node[left=of img, node distance=0cm, rotate=90, anchor=center,yshift=-1.5cm,font=\color{black}] {SINR(dB)};
 \end{tikzpicture}
 \subcaption{SINR}
  \label{fig:23all}
\end{minipage}%
\begin{minipage}{0.5\textwidth}
\begin{tikzpicture}
  \node (img)  {\includegraphics[width=0.95\linewidth]{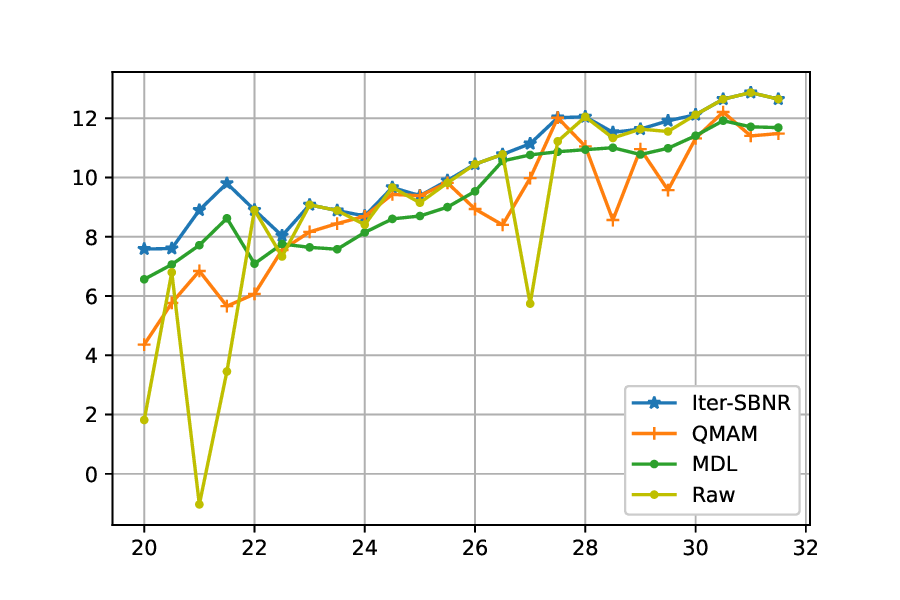}};
  \node[below=of img, node distance=0cm,xshift = 0.5cm, yshift=1.5cm,font=\color{black}] {Frequency(MHz)};
  \node[left=of img, node distance=0cm, rotate=90, anchor=center,yshift=-1.5cm,font=\color{black}] {SINR(dB)};
\end{tikzpicture}
\subcaption{SINR Zoomed}
  \label{fig:23zoomed}

\end{minipage}%
\caption{Comparison of cleaned vs. raw $\textbf{R}$ for using iterative SINR (Eq.(\ref{eq_problem_general})), MDL and  QMAM approaches. For the case of QMAM, the data used for estimating $w_{k}$ and $\epsilon$ is collected at $23h44m$ LST with RFI being removed from data 1 month apart at $23h23m$ LST.}
\label{fig:1month}
\end{figure*}
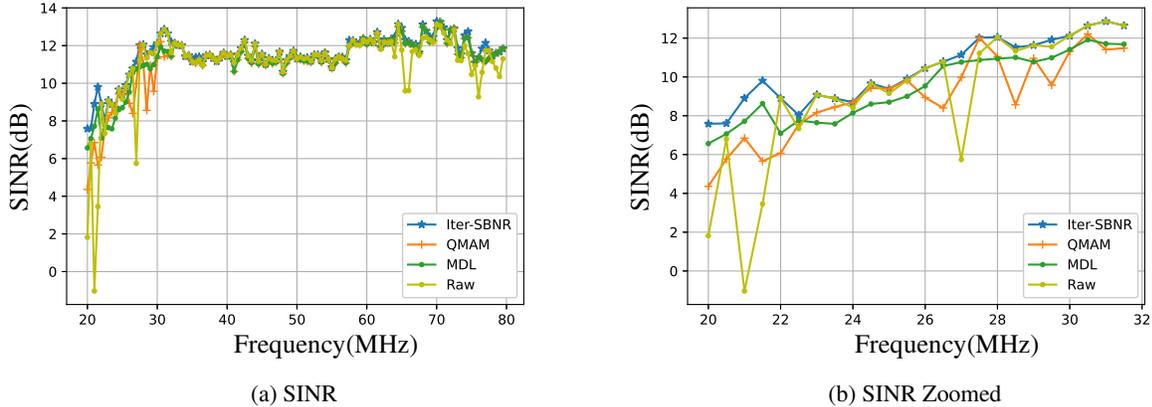

We use LWA-1 wideband data in test mode to test the proposed algorithms for RFI mitigation as mentioned in Section \ref{sec:data_processing} \cite{LWA}.  For this purpose, we demonstrate the results using the part of the spectrum contaminated with stronger RFI. Figure \ref{fig:spectrum} shows an example of the spectrum of the data from antenna stand $258$ of the LWA-1 array recorded at $0h14m$ LST. It can be seen that the region below $30\,$MHz has relatively stronger RFI.

Given the SINR measure defined above, we can also use it to compare the quality of different RFI mitigation techniques. We compare the computationally faster QMAM algorithm to the iterative-SINR approach where we find the best possible RFI subspace according to the SINR metric. We also present the results using the MDL based subspace decomposition for spatially filtering the RFI subspace. It should be noted that we have done comparisons with the standard power spectrum thresholding approach as well where we find the power per segment for the incoming signal and null down the corresponding frequency for that segment. But, given that the complete signal has RFI contamination, we either null the complete signal or keep the complete signal. Because of this reason, the results labeled as untouched/raw also represent the performance for this power-spectrum-based excision approach.

Figure \ref{fig:1day} shows the comparison of different RFI mitigation techniques for different frequency bands. We also show the results for the case where RFI mitigation is not applied to the data (raw/untouched). In this case, Cas-A is used as the source for computing SINR. For calculating the weights and the corresponding factor $\epsilon$, the data set used is from $5h11m$ LST approximately 1 day apart from $5h0m$ LST. It can be observed in Figure \ref{fig:5all} that the RFI mitigation is significantly able to improve the SINR compared to the SINR using raw signal data. As mentioned earlier, most of the RFI is contaminating the frequency band from the frequency range below 30$\,$MHz. Figure \ref{fig:5zommed} shows the zoomed results for this frequency range. It can be observed that QMAM is able to perform significantly better as compared to MDL in this case while still being computationally more efficient and reaching close to the best achievable SINR of the iterative-SINR approach. For the frequency range above 30$\,$MHz, we observe that SINRs for all cases are same since there is very little RFI in these frequency bands.

\begin{figure}[t]
\begin{center}
    
\begin{minipage}{0.7\textwidth}
\begin{tikzpicture}
  \node (img)  {\includegraphics[width=0.95\linewidth]{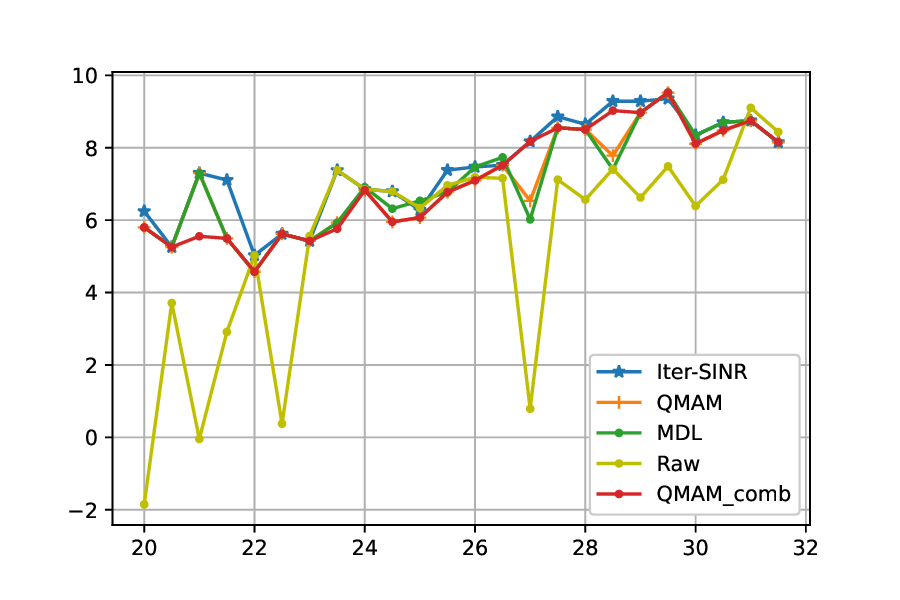}};
  \node[below=of img, node distance=0cm,xshift = 0.5cm, yshift=1.5cm,font=\color{black}] {Frequency(MHz)};
  \node[left=of img, node distance=0cm, rotate=90, anchor=center,yshift=-2cm,font=\color{black}] {SINR(dB)};
 \end{tikzpicture}
\end{minipage}%
\caption{SINR vs frequency with combined-QMAM approach. In this case, the RFI data to be mitigated is from $23h24m$ LST (UTC 04/30/2013) with data from $t_{1} = 23h14m$ LST and $t_{2} = 23h44m $ LST (UTC 03/27/2013) used to estimate $\epsilon_{f} = \frac{\epsilon_{t_{1}} + \epsilon_{t_{2}}}{2} $.}
\label{fig:QMAM_combined}
\end{center}
\end{figure}

\begin{figure}[htp]
\begin{minipage}{0.5\textwidth}
\begin{tikzpicture}
  \node (img)  {\includegraphics[width=0.95\linewidth]{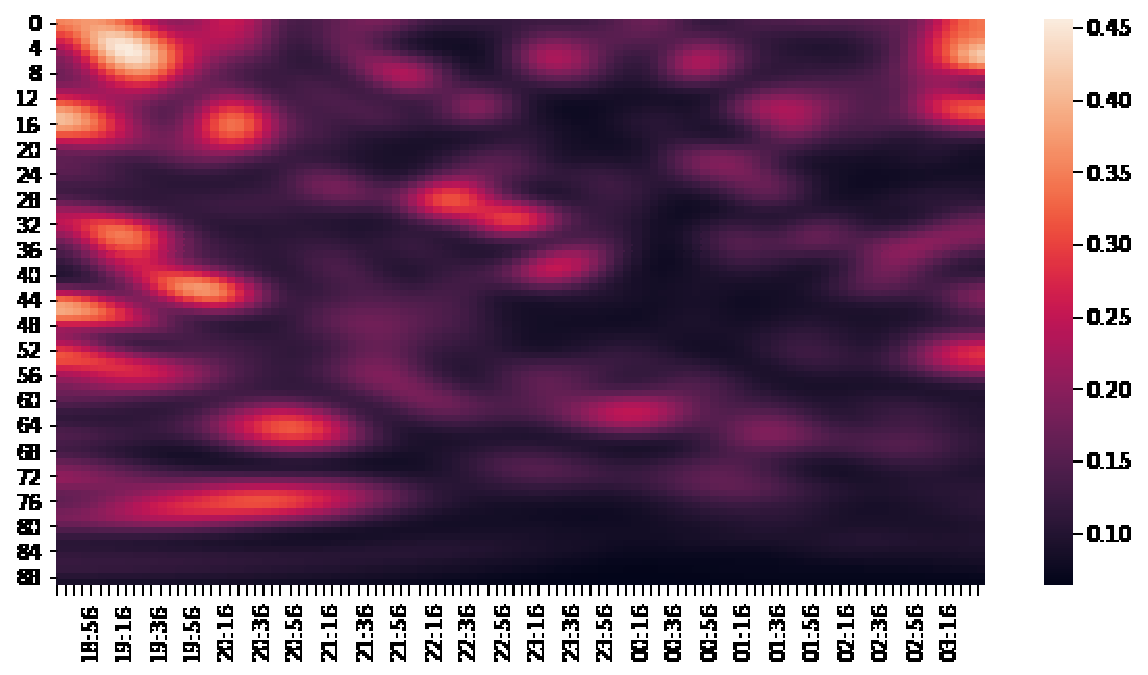}};
  \node[below=of img, node distance=0cm,xshift = 0cm, yshift=1.2cm,font=\color{black}] {Right Ascension (HH:MM)};
  \node[left=of img, node distance=0cm, rotate=90, anchor=center,yshift=-1cm,font=\color{black}] {Declination$/^{o}$};
\end{tikzpicture}
 \subcaption{RFI at LST $23h29m$, 27.5 MHz}
 \label{fig:imageRFIfree_27.5MHZ}
\end{minipage}%
\begin{minipage}{0.5\textwidth}
\begin{tikzpicture}
  \node (img)  {\includegraphics[width=0.95\linewidth]{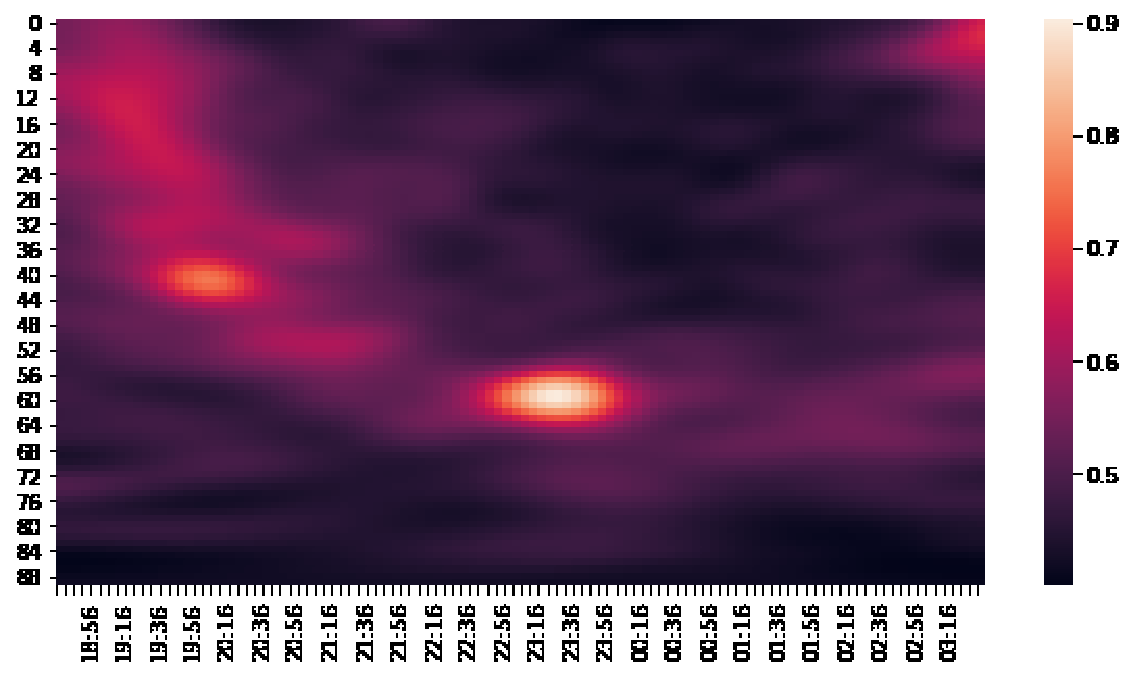}};
  \node[below=of img, node distance=0cm,xshift = 0cm, yshift=1.2cm,font=\color{black}] {Right Ascension (HH:MM)};
  \node[left=of img, node distance=0cm, rotate=90, anchor=center,yshift=-1cm,font=\color{black}] {Declination$/^{o}$};
\end{tikzpicture}
 \subcaption{RFI Mitigated at LST $23h29m$, 27.5 MHz}
 \label{fig:imageRFIfree_27.5MHZ}
\end{minipage}%

\begin{minipage}{0.5\textwidth}
\begin{tikzpicture}
  \node (img)  {\includegraphics[width=0.95\linewidth]{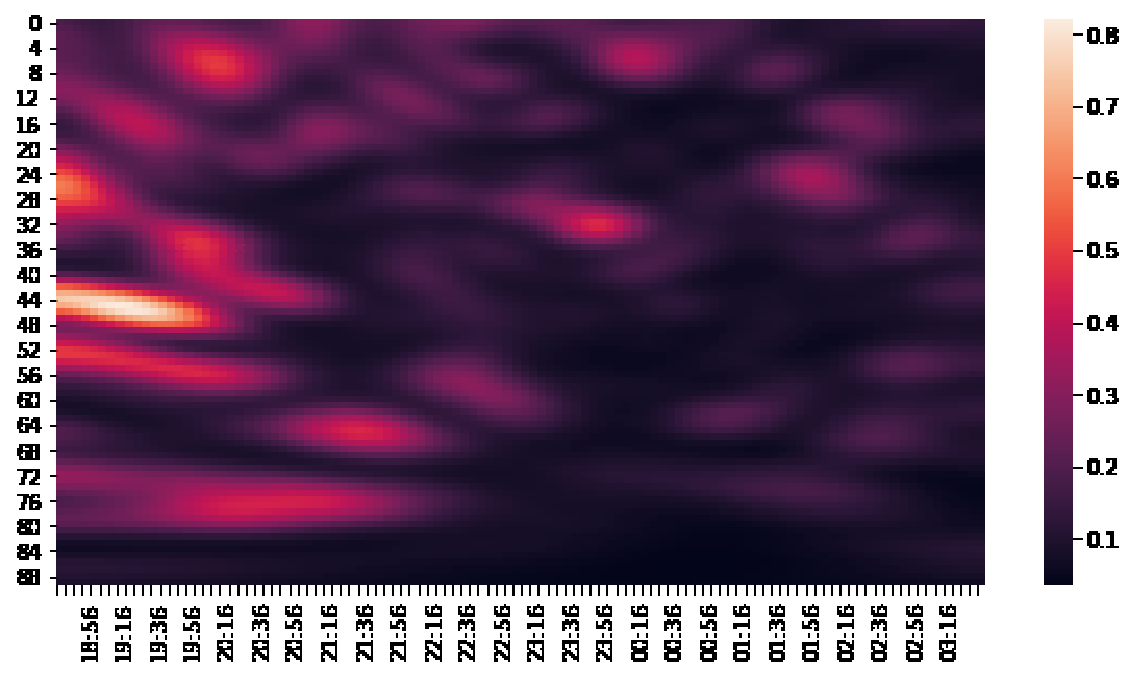}};
  \node[below=of img, node distance=0cm,xshift = 0cm, yshift=1.2cm,font=\color{black}] {Right Ascension (HH:MM)};
  \node[left=of img, node distance=0cm, rotate=90, anchor=center,yshift=-1cm,font=\color{black}] {Declination$/^{o}$};
\end{tikzpicture}
 \subcaption{RFI at LST $23h24m$, 27 MHz}
 \label{fig:imageRFIfree_27.5MHZ}
\end{minipage}%
\begin{minipage}{0.5\textwidth}
\begin{tikzpicture}
  \node (img)  {\includegraphics[width=0.95\linewidth]{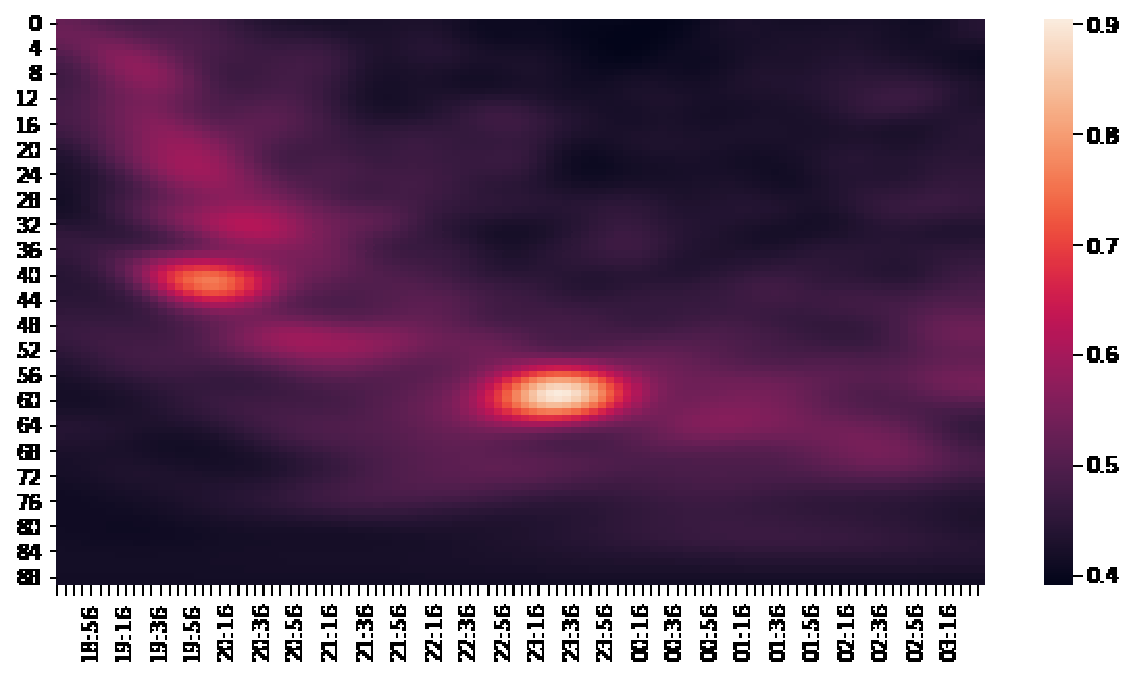}};
  \node[below=of img, node distance=0cm,xshift = 0cm, yshift=1.2cm,font=\color{black}] {Right Ascension (HH:MM)};
  \node[left=of img, node distance=0cm, rotate=90, anchor=center,yshift=-1cm,font=\color{black}] {Declination$/^{o}$};
\end{tikzpicture}
 \subcaption{RFI Mitigated at LST $23h24m$, 27 MHz}
 \label{fig:imageRFIfree_27.5MHZ}
\end{minipage}%

\begin{minipage}{0.5\textwidth}
\begin{tikzpicture}
  \node (img)  {\includegraphics[width=0.95\linewidth]{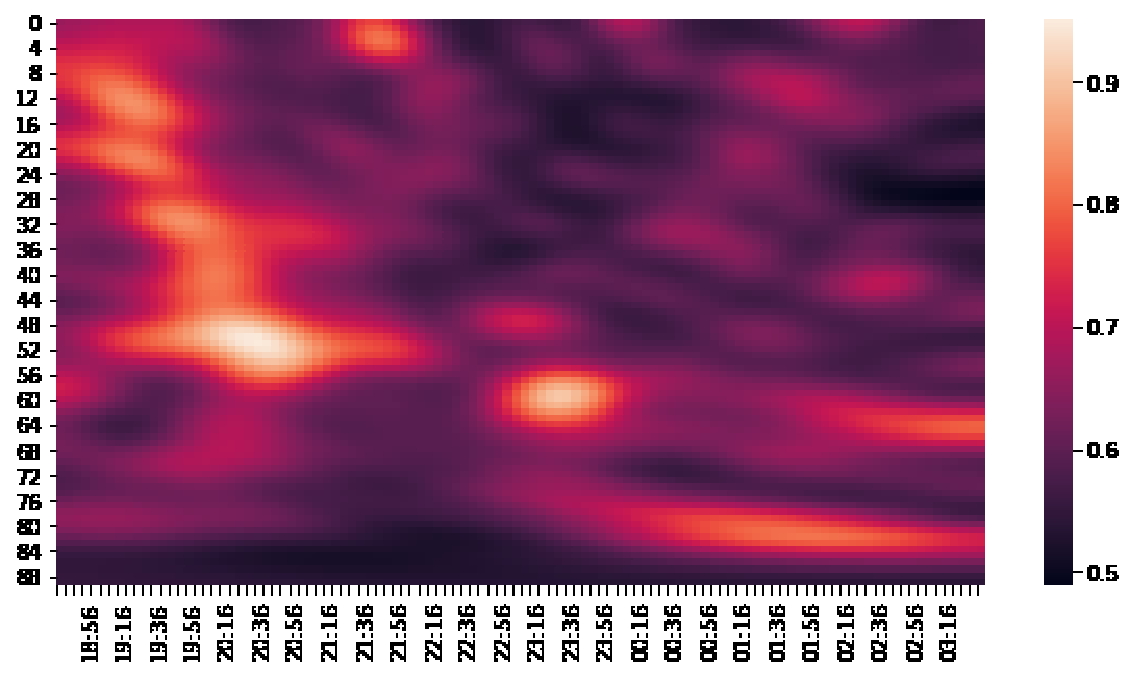}};
  \node[below=of img, node distance=0cm, xshift = 0cm,yshift=1.2cm,font=\color{black}] {Right Ascension (HH:MM)};
  \node[left=of img, node distance=0cm, rotate=90, anchor=center,yshift=-1cm,font=\color{black}] {Declination$/^{o}$ };
 \end{tikzpicture}
  \subcaption{RFI at LST $0h14m$, 27 MHz}
 \label{fig:imageRFI_27MHZ}
\end{minipage}%
\begin{minipage}{0.5\textwidth}
\begin{tikzpicture}
  \node (img)  {\includegraphics[width=0.95\linewidth]{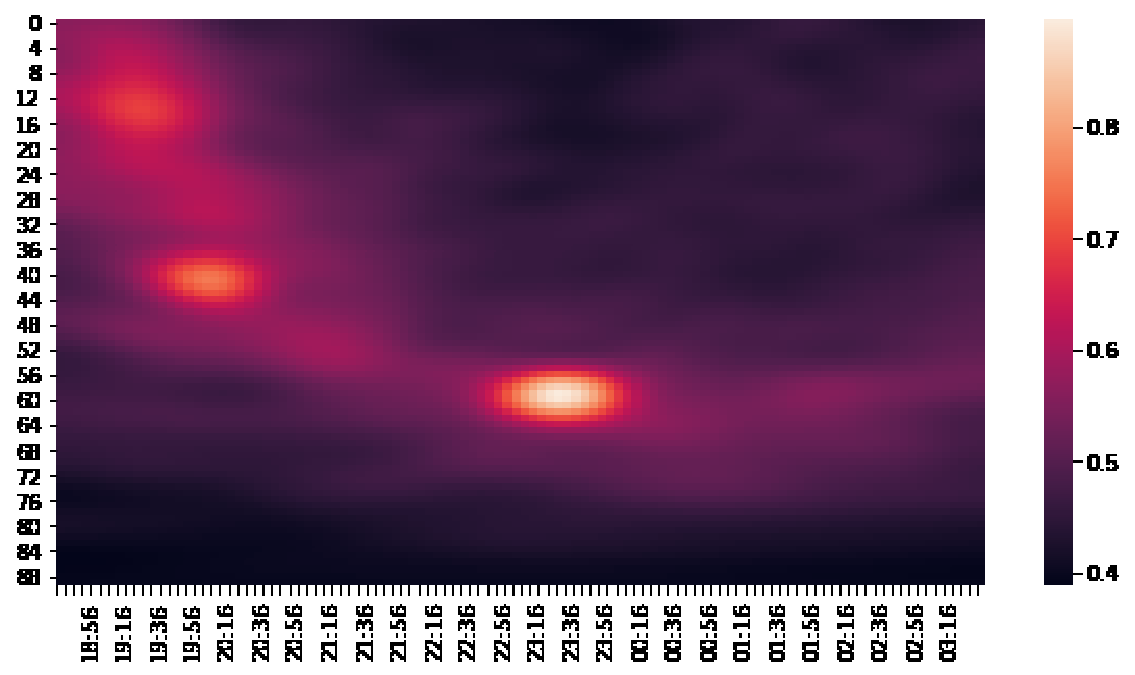}};
  \node[below=of img, node distance=0cm,xshift = 0cm, yshift=1.2cm,font=\color{black}] {Right Ascension (HH:MM)};
  \node[left=of img, node distance=0cm, rotate=90, anchor=center,yshift=-1cm,font=\color{black}] {Declination$/^{o}$};
\end{tikzpicture}
 \subcaption{RFI Mitigated at LST $0h14m$, 27 MHz}
 \label{fig:imageRFIfree_27.5MHZ}
\end{minipage}%
\caption{With and without RFI mitigation imaging pairs are shown for various frequencies and LSTs  with Cas-A (right - 58.8$^{o}$, RA: 23H23M) and Cyg-A (left - 40$^{o}$, RA: 19H59M) within the span using 60 ms of time series using the proposed method (J2000 coordinates).}
\label{fig:images}
\end{figure}

We have shown the performance of QMAM with previous day LST for estimating $\epsilon$ given by Eq. (\ref{eq:QMAM_ratio_weights}). In Figure \ref{fig:1month}, we show the comparison with QMAM using 1 month apart data under similar sky-conditions. We utilize the data from $23h44m$ LST (UTC 03/27/2013) for removing RFI from data at $23h23m$ LST (UTC 04/30/2013) with Cas-A as the source for SINR. Given the comparable results using QMAM and MDL, this further reinforces the hypothesis that the previous data under similar sky-conditions can be utilized for removing RFI.

In some cases, the fast QMAM approach still gives up performance compared to iterative-SINR and MDL. This can be attributed to slight changes in sky-conditions and  system noise for different LSTs used for approximating the RFI-free sky-conditions. As pointed out in Section \ref{sec:methodology}, we can use the combined-QMAM approach for mitigating this problem. Figure \ref{fig:QMAM_combined} shows the comparison of SINR vs. frequency with the combined-QMAM approach using different configurations of antennas. In this case, the RFI data to be mitigated is from $23h24m$ LST (UTC 04/30/2013) with data from $t_{1} = 23h14m$ LST and $t_{2} = 23h44m $ LST (UTC 03/27/2013) used to estimate $\epsilon_{f} = \frac{\epsilon_{t_{1}} + \epsilon_{t_{2}}}{2} $. Cas-A is used as a source for SINR in this case. The combined-QMAM achieves mean-squared-error (MSE) of 0.39$\,$dB$^2$ from the best possible iterative-SINR in comparison to  MDL (0.58$\,$dB$^2$), QMAM (0.49$\,$dB$^2$) and untouched/raw (10.01$\,$dB$^2$). It can be seen that for frequency 27$\,$MHz and 28.5$\,$MHz, the combined-QMAM approach is able to improve the results. Overall, the iterative-SNR is better in terms of RFI mitigation, but it is the most computationally expensive. QMAM is best in terms of computational cost and better than MDL based filtering approach in terms of RFI mitigation while achieving almost close to iterative-SINR RFI mitigation results. MDL is a blind subspace decomposition method just like QMAM. MDL uses SVD and requires $O(M^3)$ run-time complexity compared to QMAM which requires $O(dM^2)$ complexity time while still doing better than MDL.

% With and without RFI imaging for the 27$\,$MHz band with Cas-A and Cyg-A within the span using the $0h14m$ LST data and 60ms of time series using QMAM (J2000 coordinates

Given the effectiveness of the proposed approaches using the SINR measure, Figure \ref{fig:images} shows the resulting images of the sky-sources before and after RFI mitigation for bands with high RFI occupancy. The x-axis is the right ascension with the vertical y-axis being the declination at which the beam is pointed. The data from various frequency bins, LSTs and UTC (3/27/2013, 3/28/2013, 4/30/2013) is utilized for making these images. Cas-A (RA $23h23m$, declination=$58.8^{\circ}$) and Cygnus-A (RA $19h59m$, declination = $41^{\circ}$) are visible. It can be seen from Figure \ref{fig:images} that these sources are not identifiable before RFI mitigation. Cas-A is the most prominent as it is the strongest radio source at LST $0h14m$. Cygnus-A (Cyg-A) is relatively less bright but is more easier to see. Furthermore, the galactic background is also more prominent after RFI mitigation. We also look at the effect of removing the RFI on the remaining space by projecting the point spread function for a particular declination and right ascension before and after removing RFI and performing the beamforming. The resulting mean squared error,  used as a figure of merit, is 7.1e-3, and the signal to distortion ratio is = 141.7 for different frequencies contaminated with RFI (see appendix for more on signal to distortion ratio ). Thus, the proposed solutions perform effective RFI mitigation to improve the celestial source imaging.  

\section{Conclusion}
\label{sec:conclusion}
In this paper, we present two RFI mitigation techniques for radio astronomy. We investigate the use of QMAM for computationally faster RFI mitigation compared to traditional eigenvalue-based approaches. This is especially useful for radio astronomy in monitor mode. In doing so, we show that previous LST data under similar sky-conditions can be utilized for removing RFI. We also propose a celestial-source-based SINR measure for RFI evaluating the quality of achieved RFI mitigation. Furthermore, we propose an iterative-SINR approach for RFI mitigation based on the availability of a celestial radio source to be used as a reference. From the experiments utilizing the LWA-1 radio observatory data, the iterative-SINR approach achieves the best RFI mitigation performance, and the QMAM approach achieves similar performance while still being computationally faster.

\section*{Acknowledgement}
This work was supported by the National Science Foundation through Award No.1547452 and 1547443. We would also like to express our sincere appreciation to Jayson P. Van Marter for the meticulous proofreading and valuable editorial assistance in refining this research paper.

%Bibliography
\bibliographystyle{unsrt}  
\bibliography{references}  

\section*{Appendix}

\textbf{Signal-to-Distortion Ratio}

We calculated the signal-to-distortion ratio as a figure of merit in the following way. 
Given a resulting sample covariance matrix $\hat{\textbf{R}} = \textbf{U} \Lambda \textbf{U}^H$, with or without RFI mitigation, we can project the covariance matrix of point spread function at a given declination and right ascension as:

\begin{equation}
    Image(\delta_k,\gamma_l ) = \frac{1}{Tr(\textbf{a}_{\delta{k},\gamma_{l}  } \textbf{a}_{\delta{k},\gamma_{l}  }^{H})Tr(\textbf{R}_{A})}\Bigg[\textbf{a}_{\delta{k},\gamma_{l} }^{H}\textbf{U}\textbf{U}^H   \textbf{R}_{A}(\delta_{j},\gamma_{i})   \textbf{U}\textbf{U}^H  \textbf{a}_{\delta{k},\gamma_{l} }\Bigg]
\end{equation}

where the covariance matrix corresponding to the point source is $\textbf{R}_{A}(\delta_{j},\gamma_{i}) = \textbf{a}_{\delta{j},\gamma_{i} }\textbf{a}_{\delta{j},\gamma_{k} }^{H}$ placed at the declination $\delta_{j}$ and right ascension $\gamma_{i}$ representing a single direction, with $\textbf{a}_{\delta{j},\gamma_{i} }$ representing the steering vector for the given declination and right ascension. The resulting MSE, calculated over the point spread function, is for the eigenvectors $\textbf{U}$, with and without RFI subspace. The signal-to-distortion ratio is then defined as the inverse of MSE.  

\end{document}